\documentclass[seceq]{ptptex}

\usepackage{graphicx}
\usepackage{amsmath,amssymb}
\def\OneHalf {\frac{1}{2}}
\def\ThreeHalf {\frac{3}{2}}
\def\FiveHalf {\frac{5}{2}}
\def\SevenHalf {\frac{7}{2}}
\def\NineHalf {\frac{9}{2}}

\def\simge{%  ``greater than about'' symbol
    \mathrel{\rlap{\raise 0.511ex
        \hbox{$>$}}{\lower 0.511ex \hbox{$\sim$}}}}
\def\simle{%  ``less than about'' symbol
    \mathrel{\rlap{\raise 0.511ex
        \hbox{$<$}}{\lower 0.511ex \hbox{$\sim$}}}}

\title{%        %You can use \\ for explicit line-break
Strange Tribaryons as Nona-quark States%
}

\author{%       %Use \scshape  for the family name
Yuu \textsc{Maezawa}$^1$, Tetsuo \textsc{Hatsuda}$^1$ and Shoichi \textsc{Sasaki}$^{1,2}$%
}

\inst{%     %Affiliation, neglected when [addenda] or [errata]
$^1$Department of Physics, University of Tokyo, Tokyo 113-0033, Japan \\
$^2$RIKEN BNL Research Center, Brookhaven National Laboratory,
Upton, NY 11973-5000, USA 
}

\abst{%         %this abstract is neglected when [addenda] or [errata]
  Strange tribaryons
  as   nona-quark (9 quark) states   
  is studied to describe 
  the $S=-1$ resonance $S^0(3115)$  recently discovered 
   in the reaction $K^- + ~^4{\rm He} \rightarrow S^0 + p$.
  We have identified $S^0(3115)$  as a member of the 
  flavor 27-plet, in particular,  
  $(F_{\rm flavor},I_{\rm isospin},J_{\rm spin})=(27,1,\OneHalf)$ 
  or $(27,1,\ThreeHalf)$. 
  The color-magnetic interaction between quarks
   favors  small multiplets in flavor and spin, which 
    leads to a natural explanation that $I=1$ is the 
     lowest state among the $S=-1$ tribaryons with $J=\OneHalf$.
  Classification of the $S^+$-state  recently reported as well as 
    possible locations of other light strange tribaryons such as 
  $(10^*,0,\ThreeHalf)$ with $S=-1$, $(8,\OneHalf,\OneHalf)$ with $S=-2$ and 
 $(1,0,\ThreeHalf)$ with $S=-3$ are also discussed. 
 }

\begin{document}

\maketitle

\section{Introduction}

 Recently an exotic tribaryon state $S^0(3115)$ has been
 discovered  using the stopped $K^-$ absorption experiment
 at KEK-PS \cite{Suzuki:2004ep},
 \begin{eqnarray}
 K^- + ~^4 {\rm He} \rightarrow S^0 + p.
 \end{eqnarray}
 The mass of $S^0$ is about $3115\> \mathrm{MeV}$ and the decay width is 
 less than $21\> \mathrm{MeV}$. The peak 
 in the proton spectrum is over the background 
 with  a significance level, $13\> \sigma$. 
 It was also reported previously that an exotic tribaryon 
 state $S^+(3140)$ with its width less than $23\> \mathrm{MeV}$
 may be created  in the reaction
 $ K^- + ~^4 {\rm He} \rightarrow S^+ + n$. 
 Its significance is, however,  not high enough at the moment,
 $3.7\> \sigma$  \cite{Iwasaki:2003kj}.
 In Table \ref{tab:quantum}, quantum numbers of the 
 above tribaryons are summarized together with their
 hadronic and quark compositions.

 %
 % Table.1
 %
\begin{table}[b] 
%\vspace{0.5cm}
 \begin{center}
\caption{
The quantum numbers of the tribaryons. $S$, $I$, and $Q$ 
 express strangeness, isospin, and electric charge, respectively 
 \cite{Suzuki:2004ep,Iwasaki:2003kj}.
  Possible classifications of these tribaryons are shown in Fig.~\ref{fig:root}.
 }
 \renewcommand{\arraystretch}{1.1}
  \begin{tabular}{c|ccc|ccc|cc}
\hline
  & mass & width & significance & $\> S\>$  & $\> I\>$ & $\> Q\>$ & hadronic
  & quark  \\
  & (MeV)    & (MeV)  &  &  &   &  & structure &  structure \\
\hline \hline
 $S^0$  &  3115  &  $<$ 21  & 13 $\sigma$ &  $-1$  & $1$ & $0$ & $K^-pnn$ & $(3u)(5d)(1s)$ \\
\hline
 $S^+$  &  3140 &  $<$ 23  & 3.7 $\sigma$ &  $-1$  &  $0, 1$  &  $1$  & $K^-ppn$ & $(4u)(4d)(1s)$ \\
\hline
\end{tabular}
\label{tab:quantum}
 \end{center}
\end{table}

 Possible existence of the 
  tribaryon states as deeply bound kaonic nuclei was originally predicted 
  by  Akaishi and Yamazaki~\cite{Akaishi:2002bg}. 
  It is based on the assumption that
  there is a strong attraction between $\bar{K}$ and the nucleon
  in the $I=0$ channel. This leads to
  $\Lambda (1405)$  as a bound state of $\bar{K}$ and $N$ and 
  predicts even stronger bound states of $\bar{K}$ with light nuclei \cite{Dote:2003ac}.
  However,  there are several problems to be resolved
   in this approach, which include
   (i) high central density of the resulting kaonic nuclei 
  ($\sim 10 \rho_0$) which may invalidate the 
   description using hadronic degrees of freedom, and (ii)  difficulty to  
    explain that the $I=1$ state ($S^0$) lighter than the
    $S^+$ state.
   
   The purpose of this paper is to study the exotic tribaryons
   on the basis of a quark model. By doing this, we can
   have a natural solution of the problems
    (i) and (ii). 
   A key observation is that  flavor-multiplets with small dimensions have
  relatively small masses  due to color-magnetic interactions. This 
  leads  to a light $I=1$ flavor multiplet (27-plet)
   in the $S=-1$ sector.
   Moreover, we can make several predictions
   for $S=-1, -2$ and $-3$ tribaryons, which serves as a test of our quark 
    description.

\section{Classification in flavor, isospin and spin space}

 Let us first consider a system with nine quarks 
 confined in a one-body confining potential or in a bag.
 Let us further assume that all 9 quarks are in the 
  lowest angular momentum state ($l=0$) in the potential.
 Then the quark states are characterized by
   the quantum numbers in  color $SU_C(3)$, flavor $SU_F(3)$, 
 and spin $SU_J(2)$. Imposing the 
 total anti-symmetry of the 9 quarks together with the 
 total color-singletness in $SU_C(3)$, one finds possible
 irreducible representations allowed by symmetry constraints.
  This has been worked out by 
  Aerts \textit{et al.}~\cite{Aerts:1977rw} for
  various multi-quark systems, and we recapitulate the 
  results for  nona-quark (9 quark) system in Table \ref{tab:representation} 
  for strangeness $S=0,-1,-2,$ and $-3$.  

%
% Table. 2
%
\begin{table}[htb]
\begin{center}
\caption{
Allowed representations for nona-quark systems from the 
 constraints of total anti-symmetry and total color-singletness.
 For $S=-4,-5,-6$, one can access allowed representations through a useful 
 relation $(Y, F,I,J)\leftrightarrow (-Y, F^*,I,J)$. 
 }
% \begin{ruledtabular}
 \renewcommand{\arraystretch}{1.1}
  \begin{tabular}{cccl}
\hline
$S$ &$Y$  & $F$ (flavor) &  $( I , J)$=(isospin, spin) \\
\hline\hline
0 & 3  & $35^*$ & $(\OneHalf, \OneHalf)$ \\ 
    &     &  $64$     &  $(\ThreeHalf, \ThreeHalf)$ \\
\hline 
-1 &2& $10^*$ & $(0, \ThreeHalf)$ \\
     && $27$      & $(1, \OneHalf), (1, \ThreeHalf), (1, \FiveHalf)$ \\
     && $35$      & $(2, \OneHalf)$ \\
     && $35^{*}$& $(0, \OneHalf), (1, \OneHalf)$ \\
%     & $64$      & $(1, \OneHalf), (2, \ThreeHalf)$ \\
     && $64$      & $(1, \ThreeHalf), (2, \ThreeHalf)$ \\ % Corrected by SS
\hline
-2 &1& $8$        & $(\OneHalf, \OneHalf), (\OneHalf, \ThreeHalf), (\OneHalf, \FiveHalf), 
     (\OneHalf, \SevenHalf)$\\
     && $10$      & $(\ThreeHalf, \ThreeHalf)$\\
     && $10^*$ & $(\OneHalf, \ThreeHalf)$\\ 
     && $27$     &  $(\OneHalf, \OneHalf), (\OneHalf, \ThreeHalf), (\OneHalf, \FiveHalf), 
     (\ThreeHalf, \OneHalf), (\ThreeHalf, \ThreeHalf), (\ThreeHalf, \FiveHalf)$\\
     && $35$     & $(\ThreeHalf, \OneHalf), (\FiveHalf, \OneHalf)$ \\
     && $35^*$& $(\OneHalf, \OneHalf), (\ThreeHalf, \OneHalf)$ \\
     && $64$     & $(\OneHalf, \ThreeHalf), (\ThreeHalf, \ThreeHalf), (\FiveHalf, \ThreeHalf)$\\
\hline
-3 &0& $1$        &$(0, \ThreeHalf), (0, \FiveHalf), (0, \NineHalf)$ \\
     && $8$        &$(0, \OneHalf), (0, \ThreeHalf), (0, \FiveHalf), (0, \SevenHalf), 
     (1,\OneHalf), (1,\ThreeHalf), (1,\FiveHalf), (1,\SevenHalf)$\\
     && $10$      &$(1,\ThreeHalf)$\\
     && $10^*$ &$(1,\ThreeHalf)$\\ 
     && $27$     & $(0, \OneHalf), (0, \ThreeHalf), 
     (0,\FiveHalf), (1,\OneHalf), (1,\ThreeHalf), (1, \FiveHalf)
     (2, \OneHalf) ,(2, \ThreeHalf), (2,\FiveHalf)$\\
     && $35$     & $(1,\OneHalf), (2, \OneHalf)$\\
     && $35^*$& $(1, \OneHalf), (2, \OneHalf)$\\
     && $64$     & $(0, \ThreeHalf),
     (1,\ThreeHalf),(2, \ThreeHalf),(3, \ThreeHalf)$\\
\hline
 \end{tabular}
%\end{ruledtabular}
\label{tab:representation} 
\end{center}
\end{table}
%%%%%%%%%%%%%%%%%%%%%%%%%%%%%%%%%%%%%%

\begin{figure}[t]
 \begin{center}
 \includegraphics[height=6cm,clip]{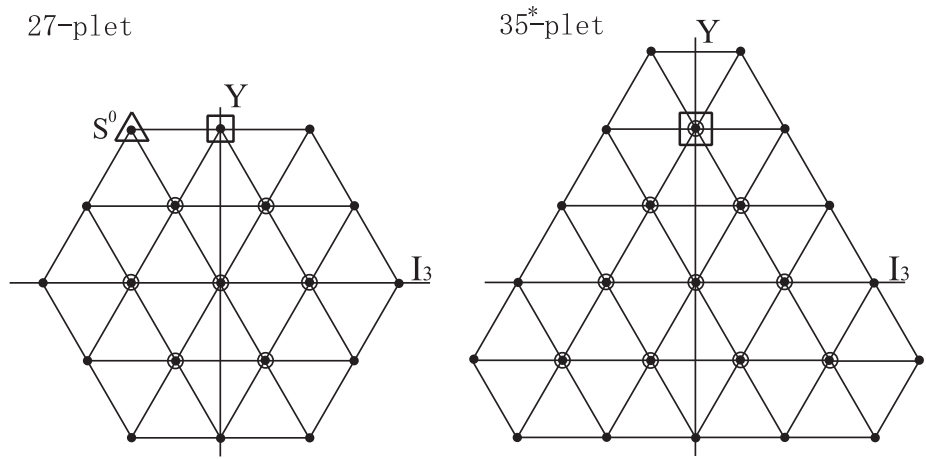}
 \caption{ The $27$-plet (left) and the $35^*$-plet (right)
  in flavor $SU_F(3)$.
  Possible location of $S^0$ ($S^+$) is shown by
 the triangle (the squares).  Note that $Y=S+3$ for the  nona-quark system.}
 \label{fig:root}
 \end{center}
 \end{figure}

 From Table \ref{tab:representation}, one finds 
  rather stringent restrictions among $F$, $I$ and $J$.
  For example, in the $S=-1$ tribaryon channels,
  the flavor 27-plet does not allow $I=0$ states.
  For later convenience, we show  the $27$-plet and 
  the  $35^*$-plet in Fig. \ref{fig:root}. As will be discussed later,
   the location 
  of a triangle is a candidate for the  $S^0$ state, while
  the squares are the possible locations of the  $S^+$ state.

%
 % Fig. 1
 %

\section{Mass formula for the tribaryon states} 

 The  states shown in Table \ref{tab:representation} 
 have mass splittings mainly due to the 
 $SU_F(3)$ breaking effect from
  the quark mass difference ($m_u \sim m_d \neq m_s$),
 and the dynamical effect from the color-magnetic interaction. 
 The former leads  to the splittings among the states with different
 strangeness, while the latter splits different flavor$\times$spin states
 through anti-symmetrization of the total wave function.
 There are also subleading effects originating from the 
  interplay between the mass effect and the color-magnetic effect.
 
 To see the above features explicitly, we adopt
 the mass formula obtained from the MIT bag model \cite{Chodos:1974je}.
 Although a specific model is taken here, qualitative aspects
 discussed below are independent of the details of the model. 
 The Hamiltonian for multi-quark system with all quarks occupying the 
  lowest angular momentum level reads \cite{Aerts:1977rw}
 \begin{eqnarray}
 H &=& a_0 + a_2Y + a_1 \left[ C_3(F) + \tfrac{1}{3}\vec{J}^2 \right]  \nonumber \\ 
   & & \ \ \ +
   a_3 \left[ \left(\vec{I}^2-\tfrac{1}{4}Y^2 \right)+ \tfrac{1}{3}\left( \vec{J}_n^2
    - \vec{J}_s^2 \right) \right]
   + a_4 \vec{J}_s^2 + a_5Y^2 .           
\label{eq:mass}
\end{eqnarray}
 Here  $Y$ ($=S+3$ for the  nona-quark system) 
  is the hypercharge and 
 $C_3(F)$ is the quadratic Casimir invariant of the $F$-multiplet in $SU_F(3)$.
 (For example, $C_3(27)=8$ and $C_3(35^*)=12$.)
  $\vec{I}$, $\vec{J}$,  $\vec{J}_n$ and $\vec{J}_s$
    are the operators for the total isospin, the total spin, the 
    total spin for non-strange quarks,
     and the total spin for strange quarks, respectively.
 
 The coefficients $a_i$ ($i=0, \cdots, 5$) are independent of the 
 quantum numbers and 
  are the functions of model parameters, such as the 
   bag pressure $B$, the current quark masses $m_{u,d,s}$, the effective
 fine-structure constant $\alpha_s^{\rm eff}$ and the bag radius $R$. 
 We assume that $m_{u,d}=0$ in the following. 
 
 The first coefficient $a_0$, which originates mainly from the 
 volume and Casimir energies of the bag and the 
 averaged kinetic energy of the 9-quarks, determines an approximate mass
  of the tribaryons. We will take this as an adjustable parameter to
   reproduce the experimental mass of  $S^0(3115)$.
  $a_2$ and $a_1$ are the major contributions to  cause mass splittings
  and are written as
\begin{eqnarray}
a_2 &=& \omega_n - \omega_s + 2 \frac{\alpha_s^{\rm eff}}{R} \left( M_{nn} - M_{ns} \right) ,
\label{eq:a1}
\\
a_1 &=& \frac{\alpha_s^{\rm eff}}{R} M_{ns}  ,
\label{eq:a2}
\end{eqnarray}
where $\omega_i  (= \left[ x_i^2 + (m_iR)^2 \right]^{1/2}/R)$ 
 is the eigenfrequency of a quark with flavor $i$ confined in the bag.
 $i=n (s)$ implies the non-strange (strange) quark.
 $x_n=2.043$ and $x_s$ is given in \cite{Chodos:1974je}.
 $M_{ij}$ are the matrix elements of the  color-magnetic interactions  
 given in the Appendix. 
  $a_2$ ($a_1$) is dominated by the effect of
   $SU_F(3)$ breaking (color-magnetic interaction).
 On the other hand, 
 $a_{3,4,5}$ are  proportional to $\alpha_s^{\rm eff}\times SU_F(3)$-breaking
 and have relatively minor contributions to the mass splittings
  in comparison to $a_{2,1}$.
  Complete but lengthy formulas for $a_{3-5}$ in the bag model are
   given in \cite{Aerts:1977rw} and will not be recapitulated here.
 
 Instead of trying to determine the bag radius $R$ by minimizing the 
 total energy of the system, we utilize an approximate scaling law
  obtained from $a_0$;
  $R_N \simeq (N/3)^{1/4} R_3$ with $R_N$ being the radius of the 
   $N$-quark bag. Taking $R_3 \simeq 1 \>$fm in the original MIT bag model,
  we estimate the size of the tribaryon bag as $R_9 \simeq 1.3 \>$fm.
  The central baryon number density of tribaryons is  
 ($\simle 5 \rho_0$) in the nona-quark description. This is a 
  number comparable to that for a single baryon and  is
   smaller than that of deeply bound kaonic nuclei by Akaishi and
    Yamazaki ($\sim 10 \rho_0$).
 For the strange quark mass, we adopt $m_s= 285\>$MeV which was
 determined to reproduce the mass splittings of octet baryons
  \cite{Chodos:1974je,Aerts:1977rw}.
 As for the effective fine-structure constant, we adopt
  two typical values in the bag model,
 $\alpha_s^{\rm eff}=1.0$ and $2.0$. The latter is close to the 
 one in the original MIT bag model. 
 
 In Table \ref{tab:coefficients}, we have shown the 
 coefficients $a_{1-5}$ for two different values of
  $\alpha_s^{\rm eff}$ with $R=1.3\ $fm.   
  Crucial observations obtained  from Eq.(\ref{eq:mass})
   together with Table \ref{tab:coefficients} are as  follows:
 \begin{enumerate}
  \item[(i)]  
   $a_2$ and $a_1$, which contain first order effects in
  $SU_F(3)$-breaking or $\alpha_s^{\rm eff}$, 
  give major contributions to the mass splittings
    although $a_3$ is not entirely negligible. 
  \item[(ii)] 
   Multiplets with small dimensions in flavor or in spin 
  have relatively small masses because of  the $a_1$-term.
  In particular, the small dimension in flavor is allowed
  for larger values of $|S|$ as can be seen from 
  Table \ref{tab:representation}. This may compensate
   the effect of $a_2$ which gives larger mass to larger $|S|$ 
    states.   
 \item[(iii)] Natural assignment of the $S^0(3115)$ state %$S^0$ state, 
  which has $S=-1$ and $I=1$, 
  is thus either $J=1/2$ or  $J=3/2$ states of the 27-plet,
 $(F,I,J)=(27, 1, \OneHalf\  {\rm or}\ \ThreeHalf)$,    
  in Table \ref{tab:representation}. This is because they belong
  to the lowest energy multiplet in the $S=-1$ and 
  $I=1$ channel.
  \end{enumerate}

\begin{table}[b]
 \caption{ The coefficients  $a_{1-5}$ in the Hamiltonian
  in the unit of MeV. $R$ is taken to be 1.3 fm.}
 \begin{center}
  \begin{tabular}{c|ccccc}
  \hline
$\alpha_s $ & $\quad a_2\quad  $ & $\quad  a_1\quad  $ & 
 $\quad  a_3\quad  $ & $\quad  a_4\quad  $ & $\quad  a_5\quad  $ \\
\hline  \hline
\ \ $1.0$ \ \ & $-152 $ & $20$ & $ 6.9$ & $ -2.3$ & $ -0.1$ \\
\ \ $2.0$ \ \ & $-128$  & $40$ & $ 14 $ & $ -4.7$ & $ -0.2$ \\
\hline
  \end{tabular}
\label{tab:coefficients}
 \end{center}
\end{table}

\section{Masses of strange tribaryons} 
 
Now we make a quantitative study of the strange tribaryons 
 on the basis of the mass formula given in Eq.(\ref{eq:mass}).
 Here, we tentatively identify 
  $S^0(3115)$ with $(27,1,\OneHalf)$ and adjust $a_0$ to reproduce
  the mass. Identifying $S^0$ with $(27,1,\ThreeHalf)$ does not
    lead to  appreciable change of the spectra.
 
\begin{figure}[t]
 \begin{center}
\includegraphics[height=10cm,clip]{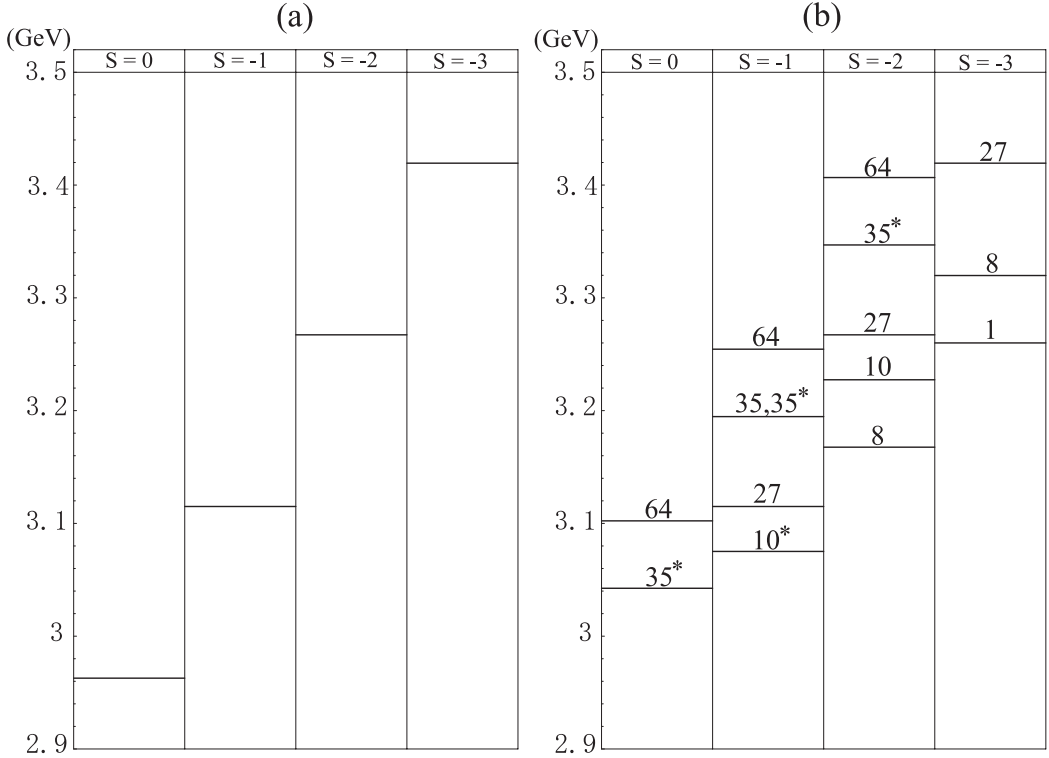}
\caption{The mass spectrum with (a) only strangeness effect (the $a_2$-term),
 and (b)  the $a_2$-term + the flavor dependent part of the $a_1$-term.}
\label{fig:spectrum1}
 \end{center}
\end{figure}

 In Fig. \ref{fig:spectrum1}(a), 
 only the $a_2$-term is taken into account 
   as an origin of the mass splittings. $a_{1}$ and $a_{3-5}$ are set to be zero.
  As expected,  adding strange quarks (increasing the hyper-charge)
  increases the mass.
  
 In Fig.\ref{fig:spectrum1}(b), we take into account the 
 flavor dependent effect of the 
  color-magnetic interaction (the term proportional to $C_3(F)$ in the $a_1$-term)
  together with $a_2$-term.   $a_{3-5}$ are still set to be zero.
 $a_0$ is readjusted so that the states in $(S,F)=(-1,27)$ 
  become $3115\>$MeV. As we have pointed out in the previous
  section, smaller (larger) flavor multiplets are relatively pushed
   down (up) in mass. Because we have neglected the spin-dependent
   term in $a_1$,  different spin states in the same multiplets
    are degenerate.
 
%
% Fig. 2
%
%
%
% Fig. 3
%
 \begin{figure}[t]
 \begin{center}
\includegraphics[width=14cm,clip]{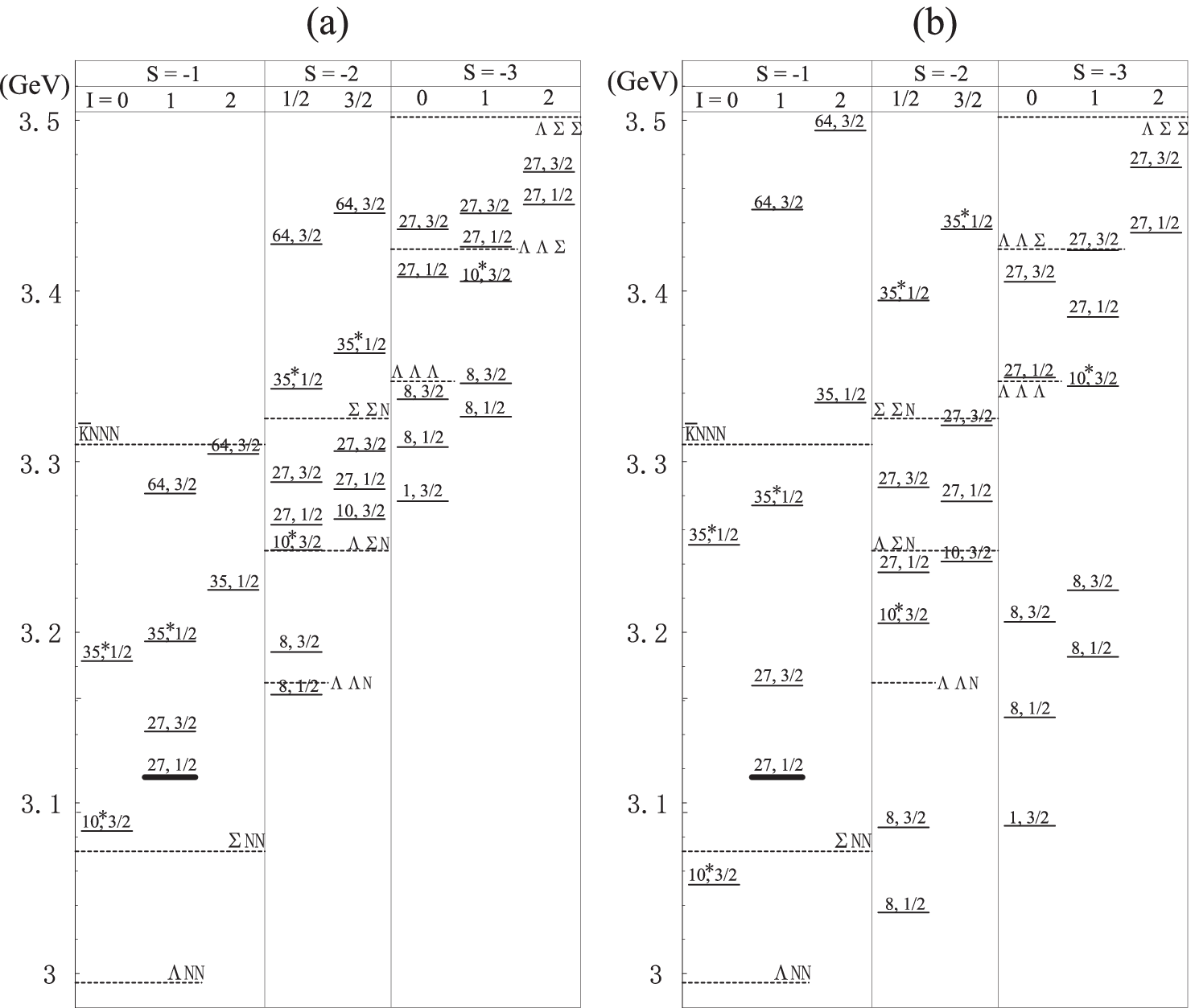}
\caption{The energy spectrum with all effects. (a) $\alpha_s^{\rm eff} =1.0$, 
 (b) $\alpha_s^{\rm eff} =2.0$ }
\label{fig:spectrum2}
 \end{center}
\end{figure}% 
  
  Finally, we show the full spectra not only with the spin-dependent part
   of the $a_1$-term but also the $a_{2,3,4,5}$-terms. 
  Fig. \ref{fig:spectrum2}(a) is the case for  $\alpha_s^{\rm eff} =1.0$ 
  and Fig. \ref{fig:spectrum2}(b) is the case for $\alpha_s^{\rm eff}=2.0$.
 We have shown only the levels with spin 1/2 and 3/2 
  and with isospin smaller than 2 
 not to make 
  the figure complicated.
   Locations of  physical thresholds to hadronic decays 
    are also indicated by the dashed lines.   
 As compared to Fig.\ref{fig:spectrum1}(b),  one can see
  spin splittings  mainly caused by
   the spin-dependent part of the color-magnetic interaction in the $a_1$-term.   
   
 In Figs.\ref{fig:spectrum1} and \ref{fig:spectrum2}, 
 we have identified  the $(F,I,J)=(27,1,\OneHalf)$ state with 
 $S^0(3115)$. 
  The mass is about 120 MeV (40 MeV) above the 
  $\Lambda NN$ ($\Sigma NN$) threshold and there is no
   selection rule to forbid the decay to these channels. Therefore,
  the small width of $S^0$ less than 21 MeV 
   may be explained only when the structure of $S^0$
  has small overlap with the hadronic final states.
  Our quark-model description in which the 9 quarks are 
   confined in a rather compact region of space, 
   could give rise to a natural explanation of the 
   small width, although further qualitative examination is
   necessary.
  Even if $S^0$ belongs to $(F,I,J)=(27,1,\ThreeHalf)$,
   the situation discussed above is unchanged. 
 
 The statistical significance of  $S^+(3140)$ is not high enough
  experimentally as shown in Table \ref{tab:quantum}. This is why we have
  not used $S^+$ as an input to determine a key parameter
  $\alpha_s^{\rm eff}$. 
   If we assume the existence of $S^+(3140)$, we have
    several possible scenarios:
  \begin{enumerate}
  \item[Case 1:]  The case where  $S^+(3140)$ belongs to $(F,I,J)=(35^*,0,\OneHalf)$.
 This naturally explains the reason why $S^0$ with $I=1$
  is lighter than $S^+$ with $I=0$.  For example, in the 
  leading order of $\alpha_s^{\rm eff}$ and the $SU_F(3)$ breaking,
   the mass splittings between
  the spin 1/2 and spin 3/2 states read
  % 
  % eq.
  % 
  \begin{eqnarray}     
  M_{35^*,0,\OneHalf}-M_{27,1,\OneHalf} &\sim & 4 M_{nn} 
  \frac{\alpha_s^{\rm eff}}{R}  >0 , \\
  M_{35^*,0,\OneHalf}-M_{27,1,\ThreeHalf} &\sim & 3 M_{nn}  
  \frac{\alpha_s^{\rm eff}}{R} >0 , 
  \end{eqnarray} 
  However,  if we try to reproduce the 25 MeV mass splitting between
  $S^+$ and $S^0$, one needs to choose
  at least 2-3 times smaller value for $\alpha_s^{\rm eff}$ 
  (or 2-3 times larger value for $R$) as compared to  the value
   usually adopted in the bag model.
 It is noteworthy here   that the $N-\Delta$ splitting
 for the  3-quark system in the bag model is also due to the color-magnetic
  interaction  and is written as 
  $M_{\Delta}-M_N= 4 M_{nn} \alpha_s^{\rm eff}/{R_3}$ 
  which is as large as  300 MeV. 
   \item[Case 2:] The case where  $S^+(3140)$ belongs to $(F,I,J)=(35^*,1,\frac{1}{2})$.
  The situation is similar to the previous case.
 \item[Case 3:] The case where  $S^+(3140)$ belongs to $(F,I,J)=(27,1,\frac{1}{2})$.
 In this case,  $S^+$ is an isospin partner of $S^0$ and  
  the 25 MeV splitting between  the two  must originate 
  from the  isospin-breaking effect. However, natural
   isospin splitting is an order of magnitude smaller, e.g.
    $M_{\Sigma^0} - M_{\Sigma^+} \simeq 3\>$MeV and $M_{\rm neutron}
    -M_{\rm proton} \simeq 1.3\> $MeV. 
 \item[Case 4:] The case where  $S^+(3140)$ belongs to 
 $(F,I,J)=(27,1,\ThreeHalf)$.
 In this case, the 25 MeV splitting can be naturally explained as a 
  result of the color-magnetic spin splittings as can be seen from 
    Fig.\ref{fig:spectrum2}(a).  
 \end{enumerate} 

 We need further experimental information, in particular, the spins 
  of $S^0$ and $S^+$, to make precise identification of 
   the multiplets they belong.

 Let us turn to some predictions which may serve to test the validity
 of our description of the tribaryon state.
 From the Hamiltonian Eq.(\ref{eq:mass}), the lightest $S=-1$ state must be 
 in the $(10^*,0,\ThreeHalf)$ multiplet. 
 Fig.\ref{fig:spectrum2} show that the location of this
  state may be just above (for $\alpha_s^{\rm eff}=1.0$) or 
 below (for $\alpha_s^{\rm eff}=2.0$) 
 the $\Sigma NN$ threshold. Systematic experimental search for the 
 states with spin 3/2 will shed more light the spin splittings 
  due to color-magnetic interaction.
 
  The mass formula also predicts light states in  larger $|S|$
 and  smaller isospin channels.
   %Another interesting state is the $(8,\frac{1}{2},\frac{1}{2})$ 
  For example, in the $S=-2$ sector, 
  the color-magnetic effect largely compensates 
   the $SU_F(3)$ breaking effect in $(8,\frac{1}{2},\frac{1}{2})$.
    Then it becomes a bound state which is lighter than 
    the $\Lambda \Lambda N$ threshold even for $\alpha_s =1.0$
    as shown in Fig.\ref{fig:spectrum2}(a). 

  Flavor singlet state in the $S=-3$ sector such as 
  $(F,I,J)=(1,0,\ThreeHalf)$ (the $H$ tribaryon) is also unique in the sense that
  it can be a bound state below the $\Lambda\Lambda\Lambda$ threshold
   by $70\>$MeV ($260\>$MeV) for $\alpha_s^{\rm eff}$=1 ($\alpha_s^{\rm eff}$=2).
  Analogous state in the 6-quark system is  
  the $H$ dibaryon \cite{Jaffe:1976yi}.~\footnote{
  %%%%%% FOOTNOTE%%%%%%%%%%%%%%%%%%%%%%%%%%%%%%
  Note that the  deeply bound  $H$ dibaryon  has been ruled out by the experimental
  discoveries of  double hypernuclei \cite{Sakai:1999qm}.}
  %%%%%%%%%%%%%%%%%%%%%%%%%%%%%%%%%%%%%%%%%%
  Predicting the masses of 
  multi-quark systems is always 
  difficult in any phenomenological quark models. In the present case,
  the mass of the $H$ tribaryon 
  is predicted relative to the $S^0(3115)$ state and thus is
  less ambiguous.
 
 \section{Relation to other approaches}

  First we discuss  the Skyrmion description of multi-baryon systems 
  \cite{Schat:2000dg,Schvellinger:1998sc},
   since a common concept of the $SU_F(3)$ symmetry 
   and its breaking are shared with our quark descriptions. 
 In {\it the rigid rotator approach} of the 
  $SU_F(3)$ skyrmion~ \cite{Yabu:1987hm}, 
  the lowest dimensional $SU_F(3)$ irreducible representation
   for  non-strange and strange tribaryons
   is shown to be the  $35^*$ multiplet~\cite{Schat:2000dg}.
   This is different from our quark description where smaller 
   representations such as $1$, $8$, $10$, $10^*$ and $27$
    are allowed for strange tribaryons. 
   
   An alternative way to analyze the 3-flavor Skyrme model is 
   {\it the bound state approach}~\cite{Callan:1985hy},  in which 
   the kaon is bound to (multi-)soliton solution~\cite{Schvellinger:1998sc}. 
   This  shares 
   common physics with the approach of the 
   deeply bound kaonic 
   nuclei~($\bar{K}$+nucleus states)~\cite{Akaishi:2002bg,Dote:2003ac}. 
   In particular, in both approaches,  (i) the $\Lambda(1405)$ state is well reproduced as a
     ${\bar K}+N$ bound state~\cite{Callan:1987xt}, and
      (ii) the binding energy of the kaon in the $S=-1$ 
     tribaryon is as large as $O(100)$ MeV~\cite{Schvellinger:1998sc}.  
   Therefore, one may be able to have closer comparison between our
   quark description and the Akaishi-Yamazaki's 
   hadronic description through the aid of the Skyrmion picture.
  
  It has been emphasized that the diquark correlations are
  important ingredients in understanding the multi-quark 
  systems~\cite{Anselmino:1992vg}, in particular  
  the pentaquark baryons \cite{Jaffe:2003sg}.
  Although a thorough study along this direction
  is beyond the scope of this paper, we briefly touch upon the
  group theoretical aspect of the diquark correlation for tribaryons.
  In the diquark hypothesis,  a  quark pair in the 
  flavor and color anti-symmetric channel is regarded as a
  diquark cluster in $3^*$ representation in color and flavor.
  Then the %9 quark 
   nona-quark system composed of 4 diquarks and an extra quark
  has decomposition to the irreducible flavor representation  as  
  \begin{eqnarray}
  3^* \otimes 3^* \otimes 3^* \otimes 3^*  \otimes 3  
   = 1 (3) \oplus 8 (8) \oplus 10 (2) \oplus 10^{*}(4) \oplus 27 (3)
  \oplus 35^{*} ,
  \end{eqnarray}
 where the numbers in parentheses denote the
  degeneracy in each multiplet.
 It turns out that the 35-plet and the 64-plet are not allowed in the
  diquark construction of the tribaryons. This fact together with 
  Table \ref{tab:representation} implies that the highest
  isospin states of strange tribaryons, such as $I=2$ for $S=-1$,
   $I=5/2$ for $S=-2$ and $I=3$ for $S=-3$,
    are disfavored by the diquark correlation.
  We have similar case for pentaquark baryons where
  large isospin states are excluded
  for the exotic $S=+1$ state.~\footnote{
  %%%%%%FOOTNOTE%%%%%%%%%%%%%%%%%%%%%%%%%%
   For pentaquark baryons, simple quark models
   show the decomposition, 
   $3 \otimes 3 \otimes 3 \otimes 3 \otimes 3^* =
  1 (3) \oplus 8 (8) \oplus 10 (4) \oplus 10^{*}(2) \oplus 27 (3)
  %\oplus 35^{*}$,
  \oplus 35$, %CORRECTED BY SS  
  while the diquark picture gives
  $3^* \otimes 3^* \otimes 3^* = 1 \oplus 8(2) \oplus 10^{*}$.
  Therefore, $10$-plet,  $27$-plet and $35$-plet are not 
  allowed  in the diquark picture. Thus 
  the exotic $S=+1$ state is uniquely assigned to the anti-decuplet
 in which  the $S=+1$ state is isospin singlet.}
 %%%%%%%%%%%%%%%%%%%%%%%%%%%%%%%%%%%%%%
  Spectra and mass splittings of each multiplet in the above
  require dynamical models of diquarks and their interactions.
 
  Finally, instanton-induced interactions among quarks
   which have not been taken into account in our simple
   description may have relevance to study the spectra
    of the multi-quark system. In fact, it has been pointed out 
  that it has a repulsive effect (opposite to
   the color magnetic interaction) to the $H$-dibaryon 
   and an attractive effect to the $\Theta^+$ pentaquark 
   \cite{Shinozaki:2004bp}. 
  It would be quite interesting to study how the instantons  modify the spectral 
   structure discussed in this paper.

\section{Summary}

  We have studied the strange tribaryons as %9 quark 
  nona-quark states  with a compact spatial size,
  which yields the plausible value of the central density 
 $(\simle 5 \rho_0$) within the quark description. 
 Assuming that   all the 9 quarks  are in the lowest orbit
 in a one-body potential, we have identified the
   recently discovered $S^0(3115)$ state as a member of the 
  flavor 27-plet, in particular,  $(F,I,J)=(27,1,\OneHalf)$ 
  or $(F,I,J)=(27,1,\ThreeHalf)$. 
  Due to the anti-symmetrization of the 9-quark wave function,
  smaller flavor multiplets appear for larger hypercharge, e.g.
  $1$ in $S=-3$, $8$ in $S=-2$, and $10^*$ and $27$ in $S=-1$.

 The color-magnetic interaction, after the anti-symmetrization of the 
  wave function, favors 
 small multiplets in flavor and spin.
 This leads to a natural explanation that, in the $S=-1$ sector,
  $(F,I,J)=(27,1,\OneHalf)$ is the lowest mass state for $I=1$ and
   and $(F,I,J)=(10^*,0,\ThreeHalf)$ is the lowest mass state for
   $I=0$.
 
 We have also discussed  possible classification of the $S^+(3140)$ state: 
  $(35^*,0,\OneHalf)$, $(35^*,1,\OneHalf)$, 
  $(27,1,\OneHalf)$, $(27,1,\ThreeHalf)$. 
   To make a quantitative comparison, one needs more  
   experimental  information, in particular the spins
    of $S^0$ and $S^+$.  
 To check the validity of our classification and identification,
  searching the light strange tribaryons such as 
  $(10^*,0,\ThreeHalf)$ with $S=-1$, $(8,\OneHalf,\OneHalf)$ with $S=-2$ and 
 $(1,0,\ThreeHalf)$ with $S=-3$ are proposed.  We especially  call an  
 attention to the $(1,0,\ThreeHalf)$ state in the $S=-3$ channel (the H tribaryon),
  which appears as a relatively deep bound state without the details of choosing
  model parameters.
 
 Finally, we emphasize that 
 experimental and theoretical studies of strange tribaryons together 
  with other multi-quark system such as the strange dibaryons
   and pentaquarks 
   may open a new window to the physics of exotic hadrons in QCD.

%%%%%%%%%%%%%%%%%%%%%%%%%%%%%%%%%%%%%%

\section*{Acknowledgments}
 We are grateful to members of hadron physics group in Univ. of Tokyo
 for discussions. We also thank M. Iwasaki and R. Hayano for
 information and discussions and S.S. thanks K. Iida and T. Doi for helpful communications.
This work was partially supported by the Grants-in-Aid of the
Japanese Ministry of Education, Culture, Sports, Science, and Technology
(No.~15540254 and No.~15740137).

\appendix
\section{} %Empty argument \section{} yields `Appendix'. 

We define the matrix elements of  color
 magnetic interaction following \cite{Aerts:1977rw}:
\begin{eqnarray}
M_{ij} &=& 3 \frac{\mu (m_i,R) \mu (m_j,R)}{R^2} I(m_iR,m_jR), \nonumber \\
\mu (m_i,r) &=& \frac{r}{6} \frac{4 \omega_i r + 2m_i r -3}
{2  \omega_i r (\omega_i r -1) + m_ir}, \nonumber \\
I(m_iR,m_jR) &=& 1 + 2 \int_0^R \frac{dr}{r^4} \mu (m_i,r) \mu(m_j,r) ,
\nonumber
\end{eqnarray}
where $\mu(m_i,r)$ is the magnetization density of a quark with mass $m_i$ 
\cite{Chodos:1974je}. Here $M_{nn}=0.177$ in the above definition.


\begin{thebibliography}{99}
%%%%%%%%%%%%%%%%%%%%%%%%%%%%%%%%%%%%%%%%%%%%%%%%%%%%%%%%%%%%%
% Some macros are available for the bibliography:
%  o for general use
%    \JL : general journals                 \andvol : Vol (Year) Page
%  o for individual journal 
%    \AJ   : Astrophys. J.           \NC         : Nuovo Cim.
%    \ANN  : Ann. of Phys.           \NPA, \NPB  : Nucl. Phys. [A,B]
%    \CMP  : Commun. Math. Phys.     \PLA, \PLB  : Phys. Lett. [A,B]
%    \IJMP : Int. J. Mod. Phys.      \PRA - \PRE : Phys. Rev. [A-E]     
%    \JHEP : J. High Energy Phys.    \PRL        : Phys. Rev. Lett.
%    \JMP  : J. Math. Phys.          \PRP        : Phys. Rep.
%    \JP   : J. of Phys.             \PTP        : Prog. Theor. Phys.     
%    \JPSJ : J. Phys. Soc. Jpn.      \PTPS       : Prog. Theor. Phys. Suppl.
% Usage:
%  \PRD{45,1990,345}          ==> Phys.~Rev.\ \textbf{D45} (1990), 345
%  \JL{Nature,418,2002,123}   ==> Nature \textbf{418} (2002), 123
%  \andvol{B123,1995,1020}    ==> \textbf{B123} (1995), 1020
%%%%%%%%%%%%%%%%%%%%%%%%%%%%%%%%%%%%%%%%%%%%%%%%%%%%%%%%%%%%%
  
%\cite{Suzuki:2004ep}
\bibitem{Suzuki:2004ep}
T.~Suzuki {\it et al.},
%``Discovery of a strange tribaryon S0(3115) in {\rm He}-4(stopped K-, p) reaction,''
Phys.\ Lett.\ B {\bf 597} (2004) 263.
%%CITATION = PHLTA,B597,263;%%

%\cite{Iwasaki:2003kj}
\bibitem{Iwasaki:2003kj}
M.~Iwasaki {\it et al.},
%``Evidence for a strongly bound kaonic system K^-ppn in the ^4{\rm He}(stopped
%K^-,n) reaction,''
arXiv:nucl-ex/0310018.
%%CITATION = NUCL-EX 0310018;%%
 
 %%%% Akaishi-Yamazaki

%\cite{Akaishi:2002bg}
\bibitem{Akaishi:2002bg}
Y.~Akaishi and T.~Yamazaki,
%``Nuclear anti-K bound states in light nuclei,''
Phys.\ Rev.\ C {\bf 65} (2002) 044005.
%%CITATION = PHRVA,C65,044005;%%

%\cite{Dote:2003ac}
\bibitem{Dote:2003ac}
A.~Dote, H.~Horiuchi, Y.~Akaishi and T.~Yamazaki,
%``Kaonic nuclei studied based on a new framework of antisymmetric molecular
%dynamics,''
Phys.\ Rev.\ C {\bf 70} (2004) 044313.
%[arXiv:nucl-th/0309062].
%%CITATION = NUCL-TH 0309062;%%
%\cite{Dote:2002db}
%\bibitem{Dote:2002db}
A.~Dote, Y.~Akaishi, H.~Horiuchi and T.~Yamzaki,
%``High-density anti-K nuclear systems with isovector deformation,''
Phys.\ Lett.\ B {\bf 590} (2004) 51.
%[arXiv:nucl-th/0207085].
%%CITATION = NUCL-TH 0207085;%%
%\cite{Yamazaki:2003hs}
%\bibitem{Yamazaki:2003hs}
T.~Yamazaki, A.~Dote and Y.~Akaishi,
%``Invariant-mass spectroscopy for condensed single and double anti-K nuclear
%clusters to be formed as residues in relativistic heavy-ion collisions,''
Phys.\ Lett.\ B {\bf 587} (2004) 167.
%[arXiv:nucl-th/0310085].
%%CITATION = NUCL-TH 0310085;%%


%\cite{Aerts:1977rw}
\bibitem{Aerts:1977rw}
A.~T.~M.~Aerts, P.~J.~G.~Mulders and J.~J.~de Swart,
%``Multi - Baryon States In The Bag Model,''
Phys.\ Rev.\ D {\bf 17} (1978) 260.
%%CITATION = PHRVA,D17,260;%%


%\cite{Chodos:1974je}
\bibitem{Chodos:1974je}
%A.~Chodos, R.~L.~Jaffe, K.~Johnson, C.~B.~Thorn and V.~F.~Weisskopf,
%``A New Extended Model Of Hadrons,''
%Phys.\ Rev.\ D {\bf 9} (1974) 3471. \\
T.~DeGrand, R.~L.~Jaffe, K.~Johnson and J.~E.~Kiskis,
%``Masses And Other Parameters Of The Light Hadrons,''
Phys.\ Rev.\ D {\bf 12} (1975) 2060.
%%CITATION = PHRVA,D12,2060;%%


%%%% Pentaquark

%\cite{Jaffe:1976yi}
\bibitem{Jaffe:1976yi}
R.~L.~Jaffe,
%``Perhaps A Stable Dihyperon,''
Phys.\ Rev.\ Lett.\  {\bf 38} (1977) 195
[Erratum-ibid.\  {\bf 38} (1977) 617].
%%CITATION = PRLTA,38,195;%%

 
%\cite{Sakai:1999qm}
\bibitem{Sakai:1999qm}
T.~Sakai, K.~Shimizu and K.~Yazaki,
%``H dibaryon,''
Prog.\ Theor.\ Phys.\ Suppl.\  {\bf 137} (2000) 121.
%[arXiv:nucl-th/9912063].
%%CITATION = NUCL-TH 9912063;%% 
 
 
%%%% Strange Multi-Skyrmion 

%% Rigid rotator approach
\bibitem{Yabu:1987hm}
H.~Yabu and K.~Ando,
%``A New Approach To The SU(3) Skyrme Model,''
Nucl.\ Phys.\ B {\bf 301} (1988) 601.
%%CITATION = NUPHA,B301,601;%%

%\cite{Schat:2000dg}
\bibitem{Schat:2000dg}
C.~L.~Schat and N.~N.~Scoccola,
%``Multibaryons in the collective coordinate approach to the SU(3) Skyrme
%model,''
Phys.\ Rev.\ D {\bf 62} (2000) 074010.
%[arXiv:hep-ph/0003247].
%%CITATION = HEP-PH 0003247;%%

%% Bound state approach  
%\cite{Callan:1985hy}
\bibitem{Callan:1985hy}
C.~G.~Callan and I.~R.~Klebanov,
%``Bound State Approach To Strangeness In The Skyrme Model,''
Nucl.\ Phys.\ B {\bf 262} (1985) 365.
%%CITATION = NUPHA,B262,365;%%

 %\cite{Schvellinger:1998sc}
\bibitem{Schvellinger:1998sc}
M.~Schvellinger and N.~N.~Scoccola,
%``Strange multiskyrmions,''
Phys.\ Lett.\ B {\bf 430} (1998) 32.
%[arXiv:hep-ph/9801347].
%%CITATION = HEP-PH 9801347;%%

\bibitem{Callan:1987xt}
C.~G.~Callan, K.~Hornbostel and I.~R.~Klebanov,
%``Baryon Masses In The Bound State Approach To Strangeness In The Skyrme
%Model,''
Phys.\ Lett.\ B {\bf 202} (1988) 269.
%%CITATION = PHLTA,B202,269;%%


%\cite{Anselmino:1992vg}
\bibitem{Anselmino:1992vg}
M.~Anselmino, E.~Predazzi, S.~Ekelin, S.~Fredriksson and D.~B.~Lichtenberg,
%``Diquarks,''
Rev.\ Mod.\ Phys.\  {\bf 65} (1993) 1199.
%%CITATION = RMPHA,65,1199;%%


%\cite{Jaffe:2003sg}
\bibitem{Jaffe:2003sg}
R.~L.~Jaffe and F.~Wilczek,
%``Diquarks and exotic spectroscopy,''
Phys.\ Rev.\ Lett.\  {\bf 91} (2003) 232003.
%[arXiv:hep-ph/0307341].
%%CITATION = HEP-PH 0307341;%%
 
%\cite{Shinozaki:2004bp}
\bibitem{Shinozaki:2004bp}
T.~Shinozaki, M.~Oka and S.~Takeuchi,
%``Effects of instanton induced interaction on the pentaquarks,''
arXiv:hep-ph/0409103.
%%CITATION = HEP-PH 0409103;%%
 M.~Oka,
arXiv:hep-ph/0411320.



\end{thebibliography}
\end{document}